\documentclass[aps,prb,a4paper]{revtex4}
\usepackage{hyperref}
\pdfoutput=1
\usepackage{graphicx}
\begin{document}

\begin{center}
{\Large 3D simulations of self-propelled, reconstructed jellyfish  
using vortex methods}\\
\ \\
Johannes~Toph{\o}j~Rasmussen$^a$,  Diego~Rosinelli$^b$,  Francesca~Storti$^a$,  Petros~Koumoutsakos$^b$,  Jens~Honor\'{e}~Walther$^{a,b}$
\\\ \\
$^a$ Department of Mechanical Engineering, Fluid Mechanics, DTU, Denmark\\
$^b$ Computational Science and Engineering Laboratory, ETH Zurich, Switzerland\\
\end{center}


We present simulations of the vortex dynamics associated with the self-propelled 
motion of jellyfish. The geometry is obtained from image segmentation of video 
recordings from live jellyfish\cite{Dabiri:2009b}. The numerical simulations are
performed using three-dimensional viscous, vortex particle methods with Brinkman
penalization to impose the kinematics of the jellyfish 
motion\cite{Sbalzarini:2005b,Chatelain:2008,Coquerelle:2008}. 
We study two types of strokes recorded in the experiment\cite{Dabiri:2009b}. 
The first type (stroke A) produces 
two vortex rings during the stroke: one outside the bell during the power stroke 
and one inside the bell during the recovery stroke. The second type (stroke B) produces 
three vortex rings: one ring during the power stroke and two vortex rings during 
the recovery stroke. Both strokes propel the jellyfish, with stroke B producing the
highest velocity.
The speed of the jellyfish scales with the square root of the Reynolds number. 
The simulations are visualized in a 
\href{http://hdl.handle.net/1813/14080}{fluid dynamics video}.

%

\begin{figure}[h]
   \includegraphics[width=1.00\textwidth]{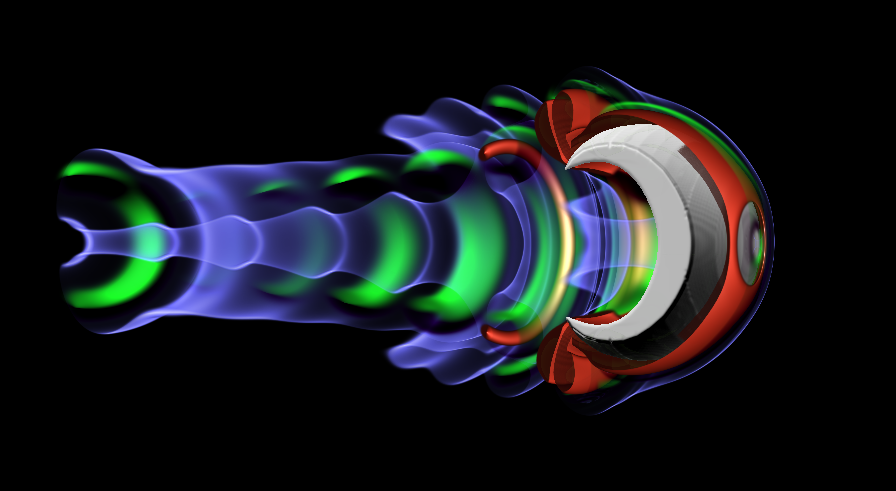}
   \caption{Visualization of the fluid vorticity (red, green, blue) of a self-propelled jellyfish. 
            The kinematics measured from
            the experiments produce three vortices during the stroke (stroke B).
            The gray isosurface illustrate the jellyfish geometry discretized using a particle level set.}
\end{figure}




\end{document}